\documentclass[twocolumn,trackchanges]{aastex631}

\usepackage{threeparttable}
\usepackage{booktabs}
\usepackage{threeparttable}
\usepackage{amsmath}
\usepackage[graphicx]{realboxes}
\usepackage{enumerate}
\usepackage{overpic}
\usepackage{xcolor} 
\usepackage{pict2e} 
\usepackage{soul}
\shorttitle{10 kpc Radio Lobes in the Sombrero galaxy}
\shortauthors{Yang Yang et al.}
\graphicspath{{./}{figures/}}

\begin{document}

\title{CHANG-ES. XXX. 10 kpc Radio Lobes in The Sombrero Galaxy}

\correspondingauthor{Yang Yang, Jiangtao Li}
\email{yyang@pmo.ac.cn, pandataotao@gmail.com}
\author[0000-0001-7254-219X]{Yang Yang}
\affiliation{Purple Mountain Observatory, Chinese Academy of Sciences, 10 Yuanhua Road, Nanjing 210023, People’s Republic of China}
\author[0000-0001-6239-3821]{Jiang-Tao Li}
\affiliation{Purple Mountain Observatory, Chinese Academy of Sciences, 10 Yuanhua Road, Nanjing 210023, People’s Republic of China}
\author[0000-0002-3502-4833]{Theresa Wiegert}
\affiliation{Instituto de Astrofísica de  Andalucía (IAA-CSIC), Glorieta de la Astronomía s/n, 18008 Granada, Spain}
\author{Zhiyuan Li}
\affiliation{School of Astronomy and Space Science, Nanjing University, Nanjing 210023, People’s Republic of China}
\affiliation{Key Laboratory of Modern Astronomy and Astrophysics, Nanjing University, Nanjing 210023, People’s Republic of China}
\author{Fulai Guo}
\affiliation{Key Laboratory for Research in Galaxies and Cosmology, Shanghai Astronomical Observatory, Chinese Academy of Sciences, 80 Nandan Road, Shanghai 200030,
People’s Republic of China}
\affiliation{University of Chinese Academy of Sciences, 19A Yuquan Road, Beijing 100049, People’s Republic of China}
\author{Judith Irwin}
\affiliation{Dept. of Physics, Eng. Phys. \& Astronomy, Queen's University, Kingston, Canada, K7L 3N6}
\author{Q. Daniel Wang}
\affiliation{Department of Astronomy, University of Massachusetts, North Pleasant Street, Amherst, MA 01003-9305, USA, LGRT-B 619E, 710}
\author{Ralf-Juergen Dettmar}
\affiliation{Ruhr University Bochum, Faculty of Physics and Astronomy, Astronomical Institute (AIRUB), 44780 Bochum, Germany}
\author{Rainer Beck}
\affiliation{Max-Planck-Institut f{\"u}r Radioastronomie, Auf dem H{\"u}gel 69, 53121, Bonn, Germany}
\author{Jayanne English}
\affiliation{Dept. of Physics \& Astronomy, University of Manitoba, Winnipeg, Manitoba, R3T 2N2}
\author{Ji Li}
\affiliation{Purple Mountain Observatory, Chinese Academy of Sciences, 10 Yuanhua Road, Nanjing 210023, People’s Republic of China}










\begin{abstract}
We report the discovery of the 10 kilo-parsec (kpc) scale radio lobes in the Sombrero galaxy (NGC~4594), using data from the Continuum Halos in Nearby Galaxies—an Expanded Very Large Array (VLA) Survey (CHANG-ES) project. We further examine the balance between the magnetic pressure inside the lobes and the thermal pressure of the ambient hot gas. At the radii $r$ of $\sim(1-10)\rm~kpc$, the magnetic pressure inside the lobes and the thermal pressure of the ambient hot gas are generally in balance. This implies that the jets could expand into the surroundings at least to $r\sim10\rm~kpc$. The feedback from the active galactic nucleus (AGN) jet responsible for the large-scale lobes may help to explain the unusually high X-ray luminosity of this massive quiescent isolated disk galaxy, although more theoretical work is needed to further examine this possibility.
\end{abstract}



\section{INTRODUCTION} \label{sec:int} 

Collimated jets and associated large-scale coherent structures carry a significant amount of energy outwards from the accreting super-massive black hole (SMBH) in the galactic nucleus. This energy is injected into the interstellar medium (ISM) or the circumgalactic medium (CGM) in the form of mechanical, thermal, cosmic ray (CR), and magnetic energy \citep{guo18,2018NatAs...2..198H,2018NatAs...2..273H,2013ARA&A..51..511K}. The X-ray bubble/cavity enveloping the radio jet is direct observational evidence of the interaction between the jet and the surrounding gas. However, most X-ray cavities are detected in massive galaxy clusters where the gas density is high \citep{2007ARA&A..45..117M}. Deep insights into an isolated \footnote{The definition of ``isolated" is based on the ``local galaxy number density" $\rho \le 0.6$.\citep{2013MNRAS.428.2085L}.} 
galaxy hosting radio jets could complement our current understanding to the physical processes of jet feedback by AGN in different environments. For example: 
How large is the sphere of influence of the jets?
Is there a balance between the magnetic pressure and the hot gas thermal pressure? 
How much mechanical energy is carried out by the jets?
Do the radio jets have enough power to heat the ambient hot gas?  

The Sombrero galaxy (NGC~4594 or M104), at a distance of $d\sim9.5$ Mpc\footnote{This distance is updated from the previous CHANG-ES standard of 12.7 Mpc  \citep{2015AJ....150...81W}.} \citep{2016AJ....152..144M}, is the most massive disk galaxy at $d\lesssim30\rm~Mpc$. The galaxy has a stellar mass of $M_{\star}\sim10^{11.3}\rm~M_{\sun}$ \citep{Kenni2011} and an unusually high rotation velocity of $v_{\rm rot}\sim379\rm~km~s^{-1}$ \citep{2023MNRAS.tmp.2366J}, inferring a total halo mass of $M_{\rm h}\sim10^{13}$~M$_{\sun}$ \citep{2020AA...643A.124K}. The galaxy is  isolated in the field with no massive companions, and the star formation (SF) is quite inactive (with a rate of ${\rm SFR}\sim0.4\rm~M_\odot~yr^{-1}$). On the other hand, its X-ray emission is unusually high compared to similarly massive or SF inactive disk galaxies \citep{2011ApJ...737...41L,2013MNRAS.428.2085L,2013MNRAS.435.3071L}. The Sombrero hosts a SMBH with a mass of $M_{\rm SMBH}\sim10^9\rm~M_{\sun}$ \citep{1988ApJ...335...40K, 1996ApJ...473L..91K} and a low Eddington ratio of $L_{\rm bol}/L_{\rm Edd}\sim10^{-5}$ \citep{2016MNRAS.459.1310K}. Existing multi-wavelength observations reveal a compact X-ray and radio core and bipolar sub-parsec (pc)- and pc-scale jets at different radio frequencies \citep{2013ApJ...779....6H, 2014ApJ...787...62M, 2006AJ....132..546G}, but there is no previous evidence of large-scale coherent structures of these radio jets.

In this paper, we present the results from our latest VLA observations of the Sombrero galaxy, which are obtained from the Continuum Halos in Nearby Galaxies—an Expanded Very Large Array Survey project (CHANG-ES, \citealt{2012AJ....144...43I,2012AJ....144...44I}).
We present our observations and data reduction in \S\ref{sec:obs}, the key results in \S\ref{sec:res}, and the scientific discussion and concluding remarks in \S\ref{sec:dis}.

\section{OBSERVATIONS AND DATA REDUCTION} \label{sec:obs} 

The VLA radio continuum data of the Sombrero galaxy (project ID:10C-119) are taken in the B, C, and D configurations in L-band (center frequency 1.5~GHz, bandwidth 512~MHz), and in C and D configurations in C-band (center frequency 6~GHz, bandwidth 2~GHz). All polarization products (Stokes I, Q, U and V) were obtained. Further details of these VLA observations are listed in Table \ref{tab:obs}.

We reduced the VLA data using the Common Astronomy Software Applications package (CASA, version 4.5) following standard procedures. Each individual visibility set was flagged, calibrated, imaged and restored. We further inspected all visibility data by eye, and manually flagged bad data (caused by radio frequency interference and instrumental effects). The Stokes I images were then produced using the \texttt{CLEAN} task, with the Multi-frequency Synthesis mode, nterms = 2, and Briggs robust 0 weighting. The CASA {\texttt{WIDEBANDPBCOR}} task was used to carry out wide-band primary beam corrections. Flux measurements were made from the primary beam-corrected images in all cases. All measured results are listed in Table \ref{tab:obs}.
The root-mean-square error (RMS) is measured in a signal-free portion (near the source) of each image. The uncertainty of the flux densities for each region was calculated using the equation
$\sigma\simeq\sqrt{N_b\times RMS^2 + (\eta\times S)^2}$, where $N_b$ corresponds to the number of synthesized beams, and $\eta$ is a factor to account for uncertainties in the calibration system, which we adopted as $\eta$ = 0.03 for the VLA radio images \citep{2013ApJS..204...19P}; $S$ is the flux density of the core, and we measured this by fitting a Gaussian to the nuclear region in each image with the \texttt{IMFIT} task. The fitting region is approximately twice the full-width-half-maximum of the synthesized beam.

Stokes Q and U maps were produced using the same sets of input parameters as the total intensity images. We derived the linearly polarized intensity image using the relation P=$\sqrt{Q^2+U^2-\sigma_{\rm Q,U}^2}$, where $\sigma_{\rm Q,U}$ is the RMS noise in the Q and U maps. 
The polarization angle of the observed electric vector ($\chi$) is given by $\chi=1/2\arctan(\rm U/Q)$; and the perpendicular direction of $\chi$ represents the apparent magnetic field orientation on the sky plane. However, we have not corrected for any potential Faraday rotation yet.

\section{RESULTS} \label{sec:res}

Our derived radio intensity map is shown in Figure \ref{fig:moph}.
In the deep VLA L-band (1.5 GHz) D-configuration total intensity image (Figure \ref{fig:moph}a), 
we for the first time detect the galactic scale bipolar radio lobes extending to $\sim10\rm~kpc$/$7\rm~kpc$ north and south of the disk. 
These FR~I-like\footnote{FR~I galaxies have a ``fan-shaped" radio morphology, with diffuse and gradually diminishing radio jets. 
FR~II galaxies have more powerful and collimated radio jets extending over greater distances, terminating in bright ``hotspots" at the outer edges of the radio lobes \citep{1974MNRAS.167P..31F}.} 
bipolar lobes are oriented almost perpendicular to the galactic disk (with P.A.\footnote{P.A. is the position angle, the direction perpendicular to the galactic disk is defined as 0$\degr$.} $\sim$ -10$\degr$). 
With the higher resolution 1.5 and 6~GHz images, we also detect some possibly coherent smaller scale (a few hundred pc to a few kpc) features of these large-scale radio lobes, with a slightly different P.A. of $\sim -20\degr$ (Figure~\ref{fig:zoomin}). 
Radio jets on a similar kpc or even smaller scales have been detected by several authors. \cite{2013ApJ...779....6H} detected the sub-pc scale bipolar jets in Sombrero at different frequencies (1.4, 2.3, 5.0, 8.4, 15.2, 23.8, and 43.2 GHz) with the VLBI, which has a P.A. of $\sim-20\degr$.
\cite{2014ApJ...787...62M} resolved a pc-scale jet into several components at 23.8~GHz with the VLBI with a P.A.$\sim-31\degr$. 
\cite{2006AJ....132..546G} detected a fainter kpc-scale (3.8~kpc) linear radio structure with archival VLA 5-GHz observations.

We then calculate the in-band spectral index $\alpha$ of the bipolar radio lobes, which is defined as $S\propto\nu^{\alpha}$ \citep{2015AJ....150...81W}. As shown in Figure~\ref{fig:moph}d, the northern (southern) lobe has an average L-band spectral index of $\alpha_{1.5}=-1.5 \pm0.9 $ ($-1.2 \pm1.0$).
For comparison, the mean spectral index of the compact core is $\sim0.3 \pm0.1$. 
Our multi-band, multi-configuration CHANG-ES data allow us to calculate the band-to-band spectral index using images with comparable angular resolution, such as $\alpha_{\rm BL-CC}$ (B-configuration L-band to C-configuration C-band spectral index). \cite{2019AJ....158...21I} reported a similarly flat radio spectrum from the core of the Sombrero galaxy ($\alpha_{\rm BL-CC}=0.43\pm0.01$, $\alpha_{\rm CL-DC}=0.07\pm0.03$), which is in general consistent with the in-band spectral index reported here. 
Because of the difficulty in subtracting the artificial effects produced by the bright central core from the extended features with the high-resolution data, we do not measure the band-to-band spectral index of the bipolar lobes. However, it is clear that there is a significant steepening of the radio spectra from the core to the larger scale structures (Figure~\ref{fig:moph}d).

Figure \ref{fig:moph}c shows the linear polarization map obtained from our D-configuration L-band data, the map is superimposed with the magnetic field orientations and Stokes~I contours. The vertically oriented cylindrical structure along the jet axis reveals the bipolar polarization lobes, which is AGN related \citep{2022ApJ...927....4Y}. 
These polarization segments are perpendicular to the bipolar polarization lobes, indicating the presence of a helical/toroidal magnetic field component traveling outward with the jet.
Additionally, the map reveals a dissociated feature with enhanced polarization emission at the endpoint of the northern lobe. At 1.5~GHz, the average fractional polarization of this feature is approximately $\sim50\%$, compared to $\sim1\%$ of the core. 
Besides the polarization core, we have not detected the bipolar polarization lobes in the better-resolution 1.5 and 6-GHz images.

\section{Discussion and Conclusions} \label{sec:dis}

\begin{figure*}[htbp]
   \includegraphics[width=.5\textwidth]{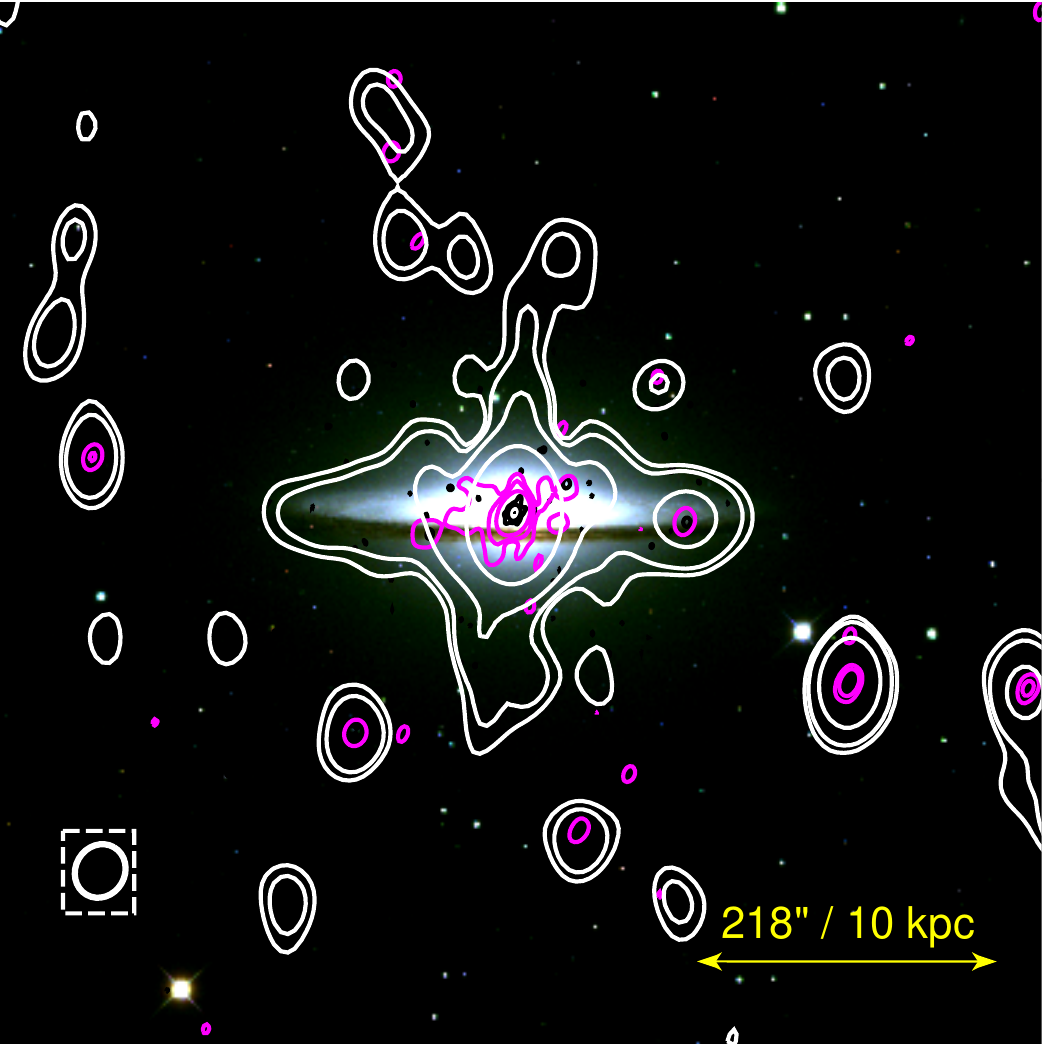} 
   \includegraphics[width=.5\textwidth,trim=26 16 18 29, clip]{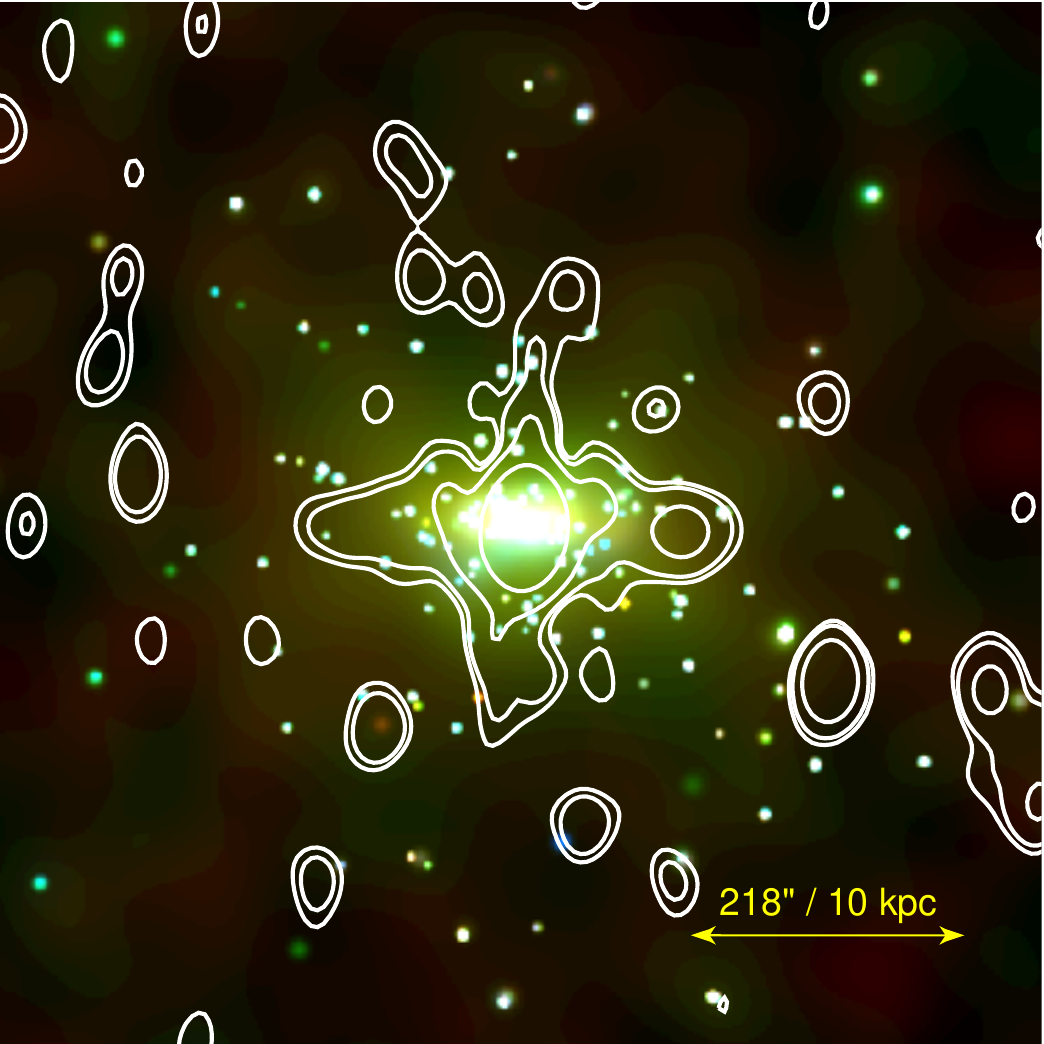}
   \includegraphics[width=.5\textwidth,trim=45 40 3 3, clip]{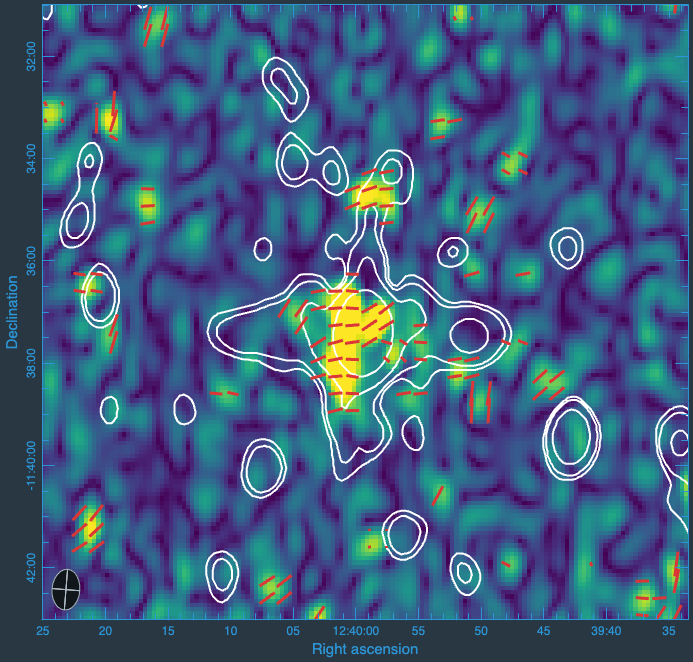}
   \begin{overpic}[width=.6\textwidth,trim=89 65 0 10, clip]{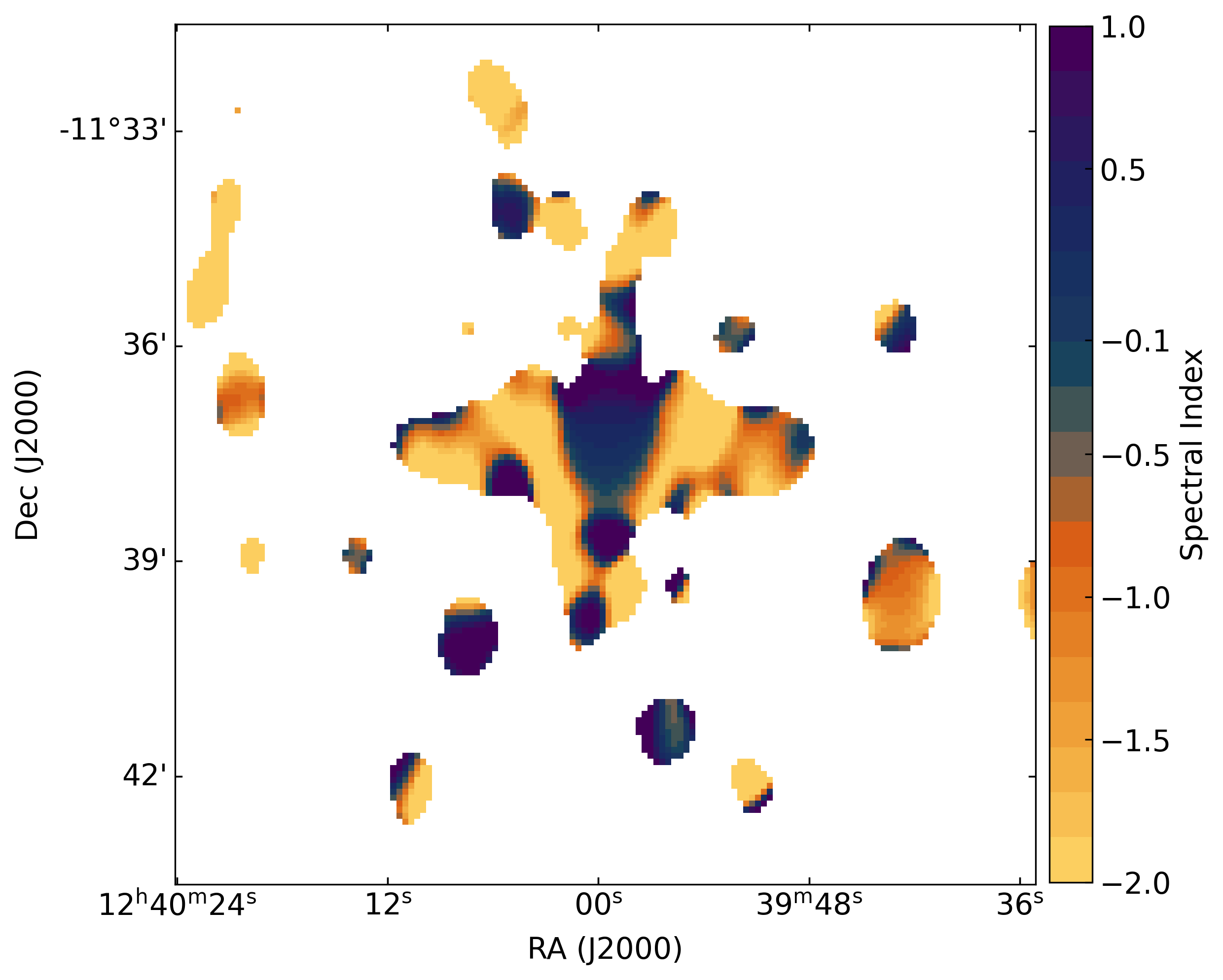}
    \put(-80,159){\color{white} \textsf{ \large a) SDSS gri images with VLA 1.5 GHz contours}}
   \put(2,159){\color{white} \textsf{\large b) Chandra X-ray image}}
   \put(-80,75){\color{white} \textsf{\large c) 1.5 GHz polarization image with B orientations}}
   \put(2,75){\color{black} \textsf{\large d) 1.5 GHz in-band spectral index}}
   \put(-35,145){\color{white} \large kpc scale}
   \put(50,155){\color{red} \large 0.5-0.8 keV}
   \put(50,150){\color{green} \large 0.8-1.5 keV}
   \put(50,145){\color{blue} \large 1.5-7.0 keV}
   \end{overpic}
   \caption{\textbf{a)} colour composite image of the global morphology of M 104, 
   created by stacking images from the Sloan Digital Sky Survey (SDSS) in the g, r, and i filters. 
   The g filter is shown in blue, while the r and i filters are shown in green and red, respectively. 
   The white VLA 1.5-GHz D-configuration contours are at levels RMS 45~$\mu$Jy~beam$^{-1}\times$[3, 5, 15, 100], 
   Zoomed-in images of the central region are shown in Figure \ref{fig:zoomin}. 
   The resolution (beam size $48\arcsec\times33\arcsec$) of the VLA data is presented in the lower left white rectangle. 
   \textbf{b)} Chandra X-ray tricolour image \citep{li2011} with the same white contours. 
   \textbf{c)} The same contours and magnetic field (B) orientations superimposed on the polarization image. The B segments have been cut off at 3$\sigma$, the RMS of the polarization image is 35 $\mu$Jy beam$^{-1}$. This image has not been corrected for Faraday rotation.
   \textbf{d)} VLA 1.5 GHz D-configuration in-band spectral index. These four images show the same field of view.}
   \label{fig:moph}
\end{figure*}

The Sombrero galaxy is unusually X-ray bright compared to other disk galaxies with comparable mass and/or SFR \citep{li2011,2011ApJ...737...41L,2013MNRAS.428.2085L,2013MNRAS.435.3071L}. The unresolved X-ray emission is significantly beyond the stellar content (Figure~\ref{fig:moph}b), indicating the presence of extended hot CGM. We then want to examine how this enhanced hot gas X-ray emission is impacting or being impacted by the jets on different scales. The first thing we want to examine is the balance between the thermal pressure of the hot gas and the pressure of the magnetic field frozen in the radio lobes, as revealed in the polarization map (Figure~\ref{fig:moph}c).

\begin{figure*}[htbp]
   \includegraphics[width=.5\textwidth]{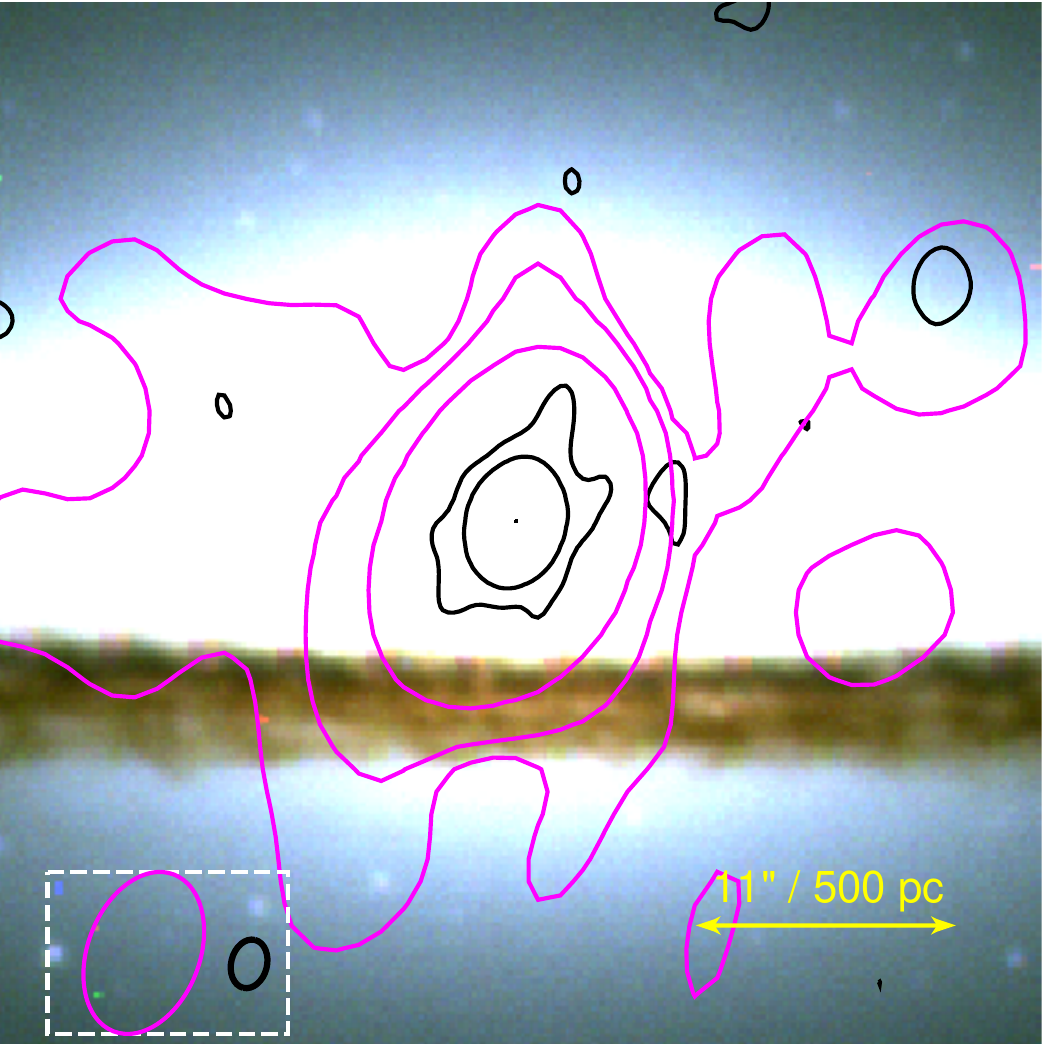} 
   \begin{overpic}[width=.5\textwidth]{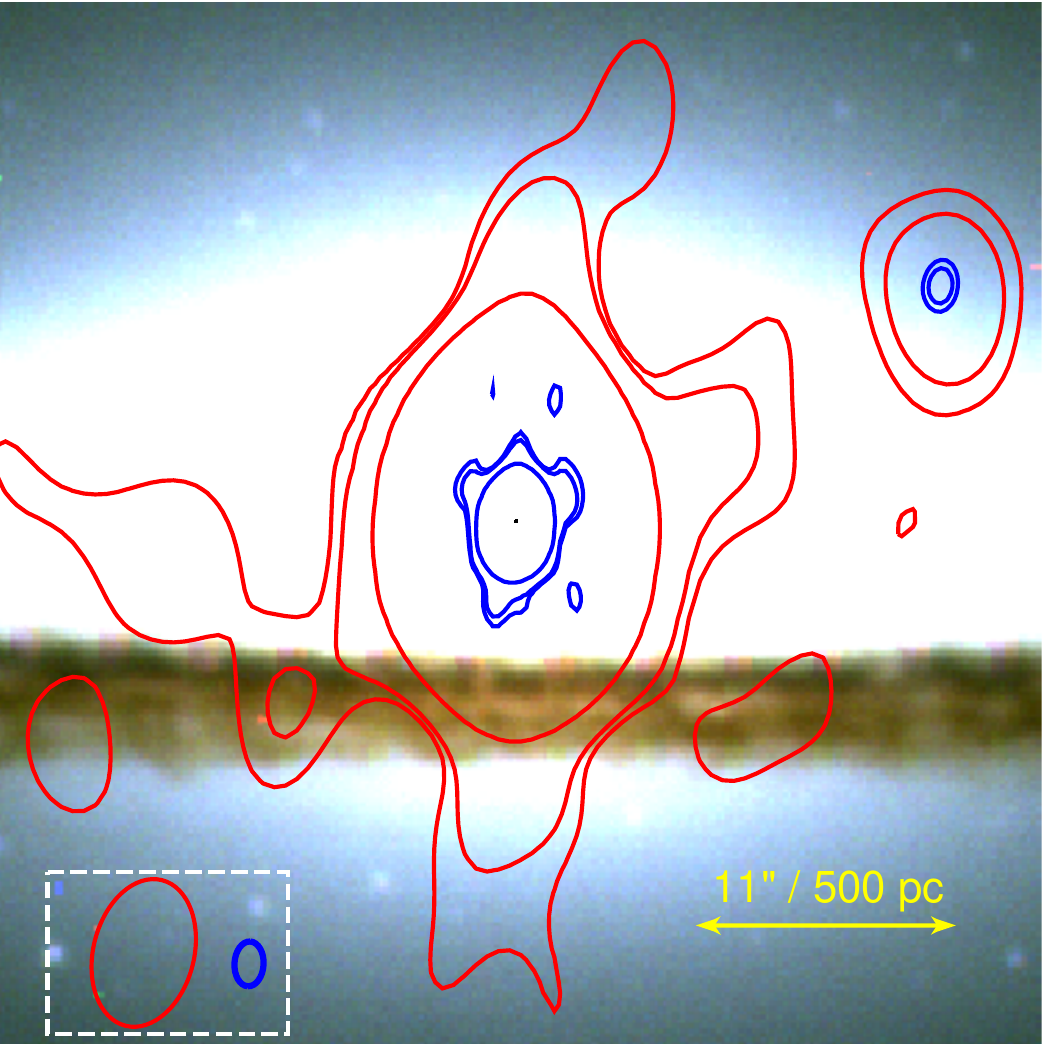}
    \put(-98,94){\color{black} \textsf{ \large a) VLA 1.5-GHz B and C configuration contours}}
   \put(2,94){\color{black} \textsf{\large b) VLA 6-GHz C and D configuration contours}}
   \put(-60,65){\color{black} \large hundreds pc}
   \put(40,65){\color{blue} \large hundreds pc}
   \put(-50,85){\color{magenta} \large sub-kpc}
   \put(35,85){\color{red} \large sub-kpc}
   \end{overpic}
   \caption{Zoomed-in images of the central region of Figure \ref{fig:moph}a. \textbf{a)} The magenta contours of the VLA 1.5-GHz C-configuration observation are at levels RMS 30~$\mu$Jy~beam$^{-1}\times$[7, 20, 100], 
   the black contours of the VLA 1.5-GHz B-configuration observation are at levels RMS 20~$\mu$Jy~beam$^{-1}\times$[3, 30]. The beam size are respectively $15\arcsec\times10\arcsec$ and $4\arcsec\times3\arcsec$ in the lower left white rectangle. 
   \textbf{b)} the red contours of the VLA 6-GHz D-configuration observation are at levels of RMS 18~$\mu$Jy~beam$^{-1}\times$[3, 5, 30], 
   the blue contours of the VLA 6-GHz C-configuration observations are at the levels RMS 5~$\mu$Jy~beam$^{-1}\times$[10, 18, 180]. The beam size are respectively $13\arcsec\times9\arcsec$ and $4\arcsec\times3\arcsec$ in the lower left white rectangle. 
   }
\label{fig:zoomin}
\end{figure*}

We estimate the magnetic field strength $B_{\rm eq}$ from the synchrotron emission of the radio lobe, based on the assumption of equipartition between the energy densities of the CRs and that of the magnetic field. This assumption is premised on an ideal scenario that CRs and magnetic fields are strongly coupled and exchange energy until equilibrium is reached over a sufficient propagation time scale and length scale (typically $\sim1\rm~kpc$; e.g., \citealt{2019Galax...7...45S}). 
We use the revised equipartition formula of \cite{2005AN....326..414B} to calculate $B_{\rm eq}$ at different places of the radio lobes. Here we assume the number density ratio of the CR proton and electron to be $K=100$, and use the corresponding in-band spectral index (in Table \ref{tab:jet}) of the hundreds-pc, sub-kpc, and kpc scale regions, as labelled in Figures~\ref{fig:moph} and \ref{fig:zoomin}. In addition, the path lengths through the emitting medium along the line of sight are assumed to be equal to the width of the extended radio structures on the sky plane. Derived $B_{\rm eq}$ values of different extended radio structures are listed in Col.~7 of Table~\ref{tab:jet}. 
The $B_{\rm eq}$ of the kpc and hundreds-pc scale regions are typically 5-7~$\mu$G and 26-51~$\mu$G, respectively. 
The number density ratio $K$ is expected to increase with increasing distance from the injection sites of CRs due to the much more severe energy losses of electrons than protons. Our measured $B_{\rm eq}$ would be underestimated by a factor of ($K_{\rm real~velue}/K$)$^{1/(3+\alpha)}$, but the equipartition magnetic pressure, $P_{B}=B_{eq}^2/8\pi$, is not very sensitive to $K$.
\cite{2006A&A...448..133K} estimated $B_{\rm eq}$ of NGC~4594, using the same method and assumptions as we used. They obtained an average $B_{\rm eq}$ of $4\pm1$~$\mu$G over the entire galaxy excluding the nucleus, which is consistent with our measurements of the kpc-scale lobes.

We also calculate the equipartition magnetic pressure $P_{B}=B_{eq}^2/8\pi$ and the synchrotron half-power lifetime of CR electrons $t_{syn}$ (using $B_{\rm eq}$ and the center frequency of the corresponding band; \citealt{2013pss5.book..641B}), which are listed in Col. 8 and 9 of Table~\ref{tab:jet}. The typical value of $t_{syn}$ of the 10~kpc-scale lobes is $\sim42-77$~Myr. This value is much larger than the typical acceleration time of CRs of a few million years.

\begin{figure*}[htbp]
   \centering
   \includegraphics[width=.6\textwidth]{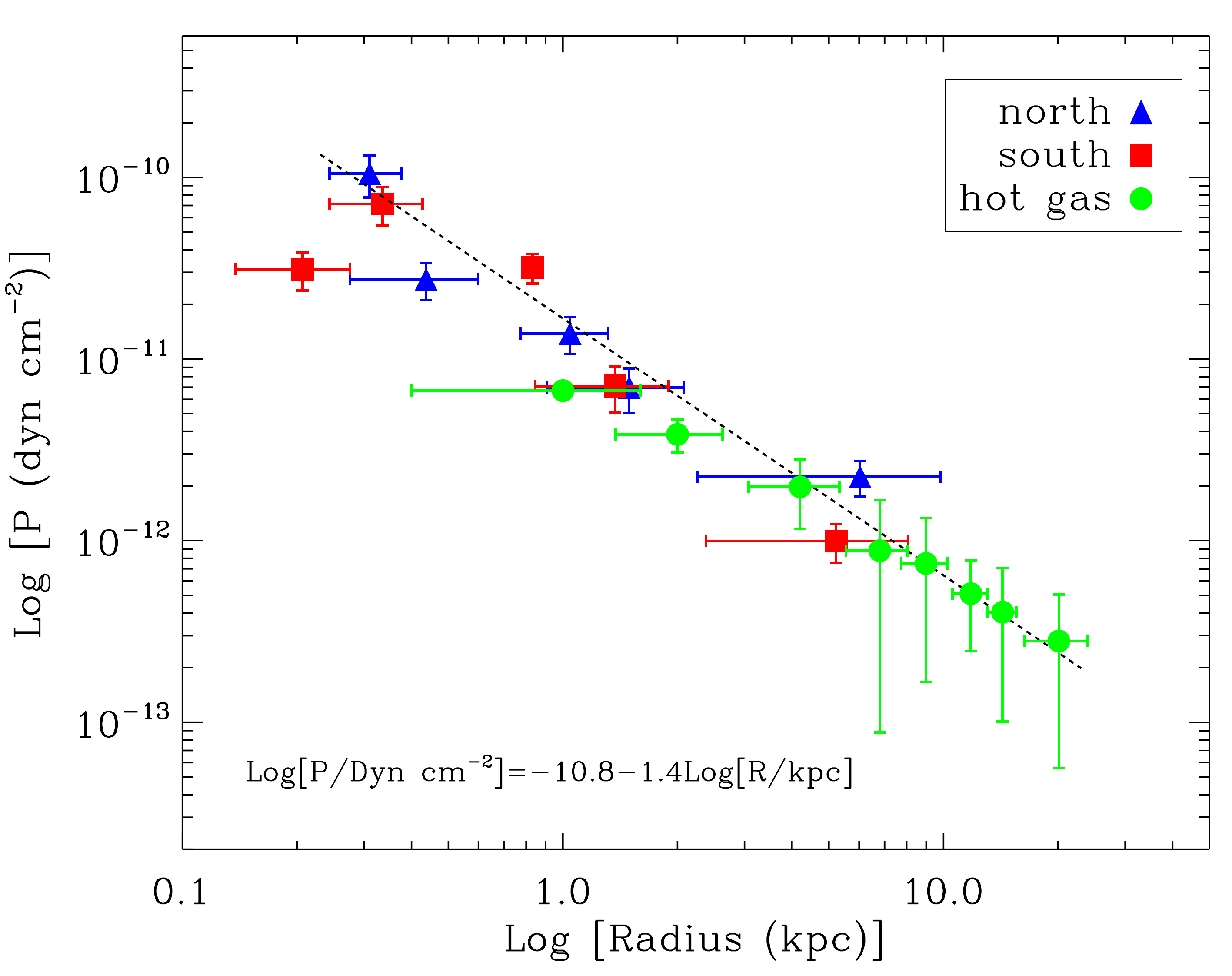} \\
   \caption{The magnetic pressure and hot gas pressure along the radio jets: The x-axis indicates the distance from the center of M104 on the sky plane. 
   The blue triangles indicate the magnetic pressure of the northern jet, the red squares are that of the southern jet, and the green circles show the profile of the hot gas pressure (hot gas data are from Fig. 9 of  \cite{li2011}), the dashed line is a linear fit of the red and blue data points with a slope of -1.4. }
   \label{fig:press}
\end{figure*}

We further compare the magnetic and hot gas pressures (derived from the Chandra observations of \citealt{li2011}) along the direction of the radio jets in Figure~\ref{fig:press}. Apparently, our measured magnetic pressure of the radio jets is comparable to the thermal pressure of the hot gas in the ambient medium. This suggests that the jets in NGC~4594 could expand into the surrounding hot medium up to a distance of at least $\sim10\rm~kpc$, mainly being driven by the magnetic pressure (or the CR pressure in balance). However, it should be noted that the real viewing angle of the large-scale jets from the line-of-sight remains uncertain. If the viewing angle is significantly smaller, for instance, $\le 25 \degr$ of the sub-pc jet as inferred from model comparisons \citep{2013ApJ...779....6H}, the potential beam effect and longer path-length would result in a lower value of $B_{\rm eq}$ and $P_{B}$, and consequently, the real radii would be longer.

We can also estimate the amount of mechanical energy required to blow out 10~kpc-scale jets in a hot medium, under the assumption of balance between the magnetic and hot gas thermal pressures at $\sim10\rm~kpc$.
We assume a cylinder shape of the radio lobe, and calculate the total energy injection (assuming all from the jets) required to balance the thermal pressure of the hot gas: $E_{jet}$ = 4$P_{hot}V$ (in analogy with the case of a bubble filled with relativistic plasma in \citealt{2006MNRAS.372...21A}), where $V$ is the volume of the radio lobe.
Assuming that the jets propagate at the sound speed ($c_s\sim 400\rm~km~s^{-1}$) of a $kT\sim0.6\rm~keV$ hot medium, the timescale to form a 10~kpc-scale lobe would be $\sim18\rm~Myr$. This dynamical timescale is significantly shorter than the synchrotron cooling timescale of the CR electrons. Therefore, the radiative loss of the CR electrons is in general energetically unimportant for the formation of the radio lobes. We can then calculate the required average jet power: $Q_{jet}=E_{jet}/t$, where $t$ is the dynamical timescale. 
Our derived $E_{\rm jet}$ is $\sim10^{55}$~ergs, and $Q_{\rm jet}$ is 2.7 $\times10^{40}$~ergs~s$^{-1}$.

As introduced above, NGC~4594 is an unusually X-ray bright SF inactive field disk galaxy, although the X-ray radiation efficiency ($\eta$, defined as the fraction of supernova (SN) energy injection released as X-ray radiation) is still far below unity \citep{2013MNRAS.428.2085L,2013MNRAS.435.3071L}. For example, it is about two orders of magnitude X-ray brighter than the similar massive early-type SF inactive isolated disk galaxy NGC~3115 (e.g., \citealt{2011ApJ...737...41L,2011ApJ...736L..23W}). The high $\eta$ of NGC~4594 thus cannot be explained with its SF activity or the large-scale environment. 

In this paper, we for the first time discover the $\sim10\rm~kpc$ scale radio lobes in NGC~4594, and link them to the smaller scale AGN jets. We further show that the jet driven outflow could be in pressure balance with the ambient hot medium at $\sim10\rm~kpc$ scale. This apparently indicates the jet could provide additional energy injection, which may contribute to the enhanced X-ray emission (the estimated jet power above, although partially from the X-ray data, is indeed one order of magnitude higher than the observed X-ray luminosity). However, caution should be made that an increased AGN jet energy injection does not necessarily lead to  increased X-ray emission. Hydrodynamic simulations indicate that AGN jets heat the ambient diffuse gas, redistribute it to larger distances, and thus tend to lower the gas density and X-ray luminosity in the long run (e.g., \citealt{guo18}; \citealt{bourne23}). However, during the early stage of a jetted AGN outburst, the jet-induced bow shock sweeps up the ambient hot gas, significantly increasing the gas density in the shock downstream and thus potentially increasing the total X-ray luminosity. This effect may be stronger in galaxies such as NGC~4594 than in more massive galaxy clusters.    

\begin{deluxetable*}{cccccccc}
\tabletypesize{\footnotesize}
\tablecaption{Information of M104 observations and Images 
 \label{tab:obs}}
\tablehead{
\colhead{Frequency} & \multicolumn{3}{c}{1.5 GHz (L band)}&\multicolumn{3}{c}{6.0 GHz (C band)}\\
\cmidrule(r{4pt}){2-4} \cmidrule(r{4pt}){5-8}
\colhead{Array}&\colhead{D$^a$}&\colhead{C$^a$}&\colhead{B$^a$}&\colhead{D$^a$}&\colhead{C$^a$}}
\startdata 
Date of observations &2011-Dec-30&2012-Mar-30&2011-Mar-17&2011-Dec-19&2012-Feb-14\\
&&&&2014-Jun-24&\\
Total bandwidth (MHz)&512&512&512&2048&2048\\
Obs. time on M104 (min)$^b$&19&41&41&73&180\\
Flux calibrator$^c$&3C286&3C286&3C286&3C286&3C286\\
Phase (secondary) calibrator$^d$&J1246-0730&J1246-0730&J1248-1959&J1246-0730&J1246-0730\\
Zero-pol calibrator$^e$&J1407+2827&J1407+2827	&J1407+2827&J1407+2827&J1407+2827\\
uv weighting$^f$&Briggs robust=0&Briggs robust=0&Briggs robust=0&Briggs robust=0&Briggs robust=0\\
\hline
\multicolumn{8}{c}{\textbf{I image}}\\
\hline
Synth. beam$^g$ ($\prime\prime$$\times$$\prime\prime$, $\circ$ )&47.9$\times$32.6, -4.6&14.9$\times$10.1, -21.5&4.4$\times$3.2, -14.0&13.3$\times$8.9, 1.1&3.9$\times$2.6, -3.3\\
RMS$^h$($\mu$Jy beam$^{-1}$)&45&30&20&18&5&&\\
$S_{core}\pm\sigma$$^i$ (mJy)&$84.1\pm0.7$&$83.1\pm0.8$&$71.6\pm0.2$&$123.7\pm0.2$&$127.2\pm0.2$\\
Fitting size$^j$&48.7$\times$33.2, 175.5&15.2$\pm$10.2, 158.6&4.4$\times$3.3, 165.6&13.4$\pm$9.0, 1.1&3.9$\pm$2.6, -3.3&& \\
\hline
\enddata
 \vspace{0.3cm}
 \tablecomments{ $^a$ project ID: 10C-119; $^b$ Total observing time on the target galaxy before flagging; $^c$ Primary flux calibrator source used to calibrate bandpass and to  determine the absolute position angle for polarization; $^d$ Secondary gain calibrator with $<3$\% amplitude closure errors expected in all array configurations and both bands; $^e$ Zero-polarization calibrator used to determine instrumental polarization leakage terms; $^f$ Robust = 0 used in CASA clean task; 
 $^g$ Synthesized beam FWHM of major and minor axis, and position angle; $^h$ root-mean-square error (RMS) manually measured in emission-free region on each image; $^i$ Flux densities of the total intensity emission in the nuclear region; $^j$ Component size determined by fitting with 2D fit tool, convolved with beam.}
\end{deluxetable*}

\begin{deluxetable*}{clccccccccccc}[b!]
\tabletypesize{\footnotesize}
\tablecaption{Measuring Results}
\tablehead{
\colhead{Config.}&
\colhead{scale}&
\colhead{l}&
\colhead{$R$}&
\colhead{$S$}&
\colhead{$\alpha_{\rm in-band}$}&
\colhead{$B_{eq}$}&
\colhead{$P_B$}&
\colhead{$t_{syn}$}&\\
&&($\arcsec$/kpc)&($\arcsec$/kpc)&(mJy)&($S\propto\nu^{\alpha}$)&($\mu$G)&(10$^{-12}$dyn cm$^{-2}$)&(Myr)\\
(1)&(2)&(3)&(4)&(5)&(6)&(7)&(8)&(9)
}
\startdata
\multicolumn{9}{c}{L band (center frequency 1.5 GHz)}\\
\hline
B&north hundreds pc  &6.0-13.0/0.32    &3.2/0.15      &$0.22\pm0.03$    &$-1.5\pm1.5$     &26 &27 & 6 \\
 &south hundreds pc  &3.0-6.0/0.14     &3.0/0.14      &$0.11\pm0.02$    &$-1.5\pm1.5$     &28 &31 & 6 \\
C&north sub-kpc &16.8-28.6/0.54   &10.8/0.50     &$0.99\pm0.04$    &$-1.4\pm1.2$     &19 &14 & 10 \\
 &south sub-kpc &17.2-19.0/0.08   &13.9/0.64     &$1.01\pm0.03$    &$-2.4\pm1.5$     &28 &32 & 6 \\
D& north kpc &49.1-213.4/7.56  &36.4/1.67     &$2.39\pm0.12$    &$-1.5\pm0.9$     &7  &2  & 42  \\
 &south kpc  &51.7-175.4/5.69  &65.0/2.99     &$1.79\pm0.13$    &$-1.2\pm1.0$     &5  &1  & 77  \\
\hline
\multicolumn{9}{c}{C band (center frequency at 6 GHz)}\\
\hline
C&north hundreds pc  &5.3-8.2/0.13     &1.7/0.08      &$0.13\pm0.01$    &$-0.9\pm1.2$     &51 &105& 1 \\
 &south hundreds pc  &5.3-9.3/0.18     &3.2/0.15      &$0.21\pm0.01$    &$-1.2\pm1.0$     &42 &71 & 1 \\
D&north sub-kpc &19.7-45.2/1.17   &10.5/0.48     &$0.28\pm0.03$    &$-0.7\pm1.1$     &13 &7 & 9 \\
 &south sub-kpc &18.4-41.2/1.05   &10.9/0.50     &$0.23\pm0.03$    &$-0.6\pm1.0$     &13 &7  & 9 \\
\enddata
\tablecomments{Column 1: VLA observing configurations;
Column 2: the scale of the extended structure;
Column 3: the closest and farthest radial distance of the radio extended structures from the center of M104 ($\arcsec$), and the length of the radio extended structures in kpc;
Column 4: the path length of the northern radio extended structure, which we assumed to be equal to its width; 
Column 5: flux density of the northern/southern radio extended structure; 
Column 6: in-band spectral index;
Column 7: equipartition magnetic field B of the radio extended structure (estimated by the revised equipartition formula of \cite{2005AN....326..414B} with the corresponding flux density and in-band spectral index; 
Column 8: magnetic pressure, $P_B$;
Column 9: the synchrotron half-power lifetime of CR electrons \citep{2013pss5.book..641B};
}
\label{tab:jet}
\end{deluxetable*}

\begin{acknowledgments}
Y.Y. acknowledges support from the National Natural Science Foundation of China (NSFC) through the grant 12203098 and the Shanghai Sailing Program (19YF1455500). Both Y.Y. and J.T.L. acknowledge the support from the NSFC through the grants 12273111 and 12321003, and also the science research grants from the China Manned Space Project. Z.L. acknowledges support by the National Key Research and Development Program of China (No. 2022YFF0503402) and the National Natural Science Foundation of China (grant 12225302). T.W. acknowledges financial support from the grant CEX2021-001131-S funded by MCIU/AEI/ 10.13039/501100011033, from the coordination of the participation in SKA-SPAIN, funded by the Ministry of Science, Innovation and Universities (MCIU). F.G. thanks the support by the Chinese Academy of Sciences under grant YSBR-061 and Shanghai Pilot Program for Basic Research - Chinese Academy of Science, Shanghai Branch (JCYJ-SHFY-2021-013). Research in this field at AIRUB is supported by Deutsche Forschungsgemeinschaft SFB1491. 
\end{acknowledgments}

%

\vspace{5mm}
\facilities{VLA(NRAO)}

\software{CARTA \citep{2021zndo...3377984C},  
          CASA \citep{2022PASP..134k4501C}, 
          }

\end{document}